%% file: scaling.tex
\documentclass[aps,pra,superscriptaddress,twocolumn]{revtex4-2}

\usepackage[utf8]{inputenc}
\usepackage{amsmath}
\usepackage{braket}
\usepackage{amsfonts}
\usepackage{amssymb}
\usepackage{lipsum}
\usepackage{dsfont}
\usepackage{stmaryrd}
\usepackage{graphicx}
\usepackage{booktabs}
\usepackage{siunitx}
\usepackage[version=4]{mhchem}  
\usepackage{hyperref}
\usepackage[capitalise]{cleveref}
\usepackage{siunitx}
\usepackage{mathtools}
\usepackage{xcolor}
\usepackage{microtype}

\input{abbreviations.tex}
\input{local.tex}

\DeclareMathOperator{\var}{var}

\DeclareMathOperator*{\argmin}{arg\,min} 
\usepackage[normalem]{ulem}
\newcommand{\stkout}[1]{\ifmmode\text{\sout{\ensuremath{#1}}}\else\sout{#1}\fi}

\usepackage[status=draft]{fixme}
\fxsetup{theme=color}

\newcommand{\auchem}{Department of Chemistry, Aarhus University, DK-8000 Aarhus C, Denmark}
\newcommand{\auphys}{Department of Physics and Astronomy, Aarhus University, DK-8000 Aarhus C, Denmark}
\newcommand{\kvantify}{Kvantify Aps, DK-2300 Copenhagen S, Denmark}

\begin{document}
    \title{Optimizing the number of measurements for vibrational structure on quantum computers: coordinates and measurement schemes}

    \author{Marco Majland}
    \affiliation{\kvantify}
    \affiliation{\auphys}
    \affiliation{\auchem}

    \author{Rasmus Berg Jensen}
    \affiliation{\auphys}
    \affiliation{\auchem}

    \author{Mads Greisen Højlund}
    \affiliation{\auchem}

    \author{Nikolaj Thomas Zinner}
    \affiliation{\kvantify}
    \affiliation{\auphys}

    \author{Ove Christiansen}
    \affiliation{\kvantify}
    \affiliation{\auchem}

    \begin{abstract}
        One of the primary challenges prohibiting demonstrations of practical quantum advantages for near-term devices
        amounts to excessive measurement overheads for estimating relevant physical quantities such as ground state
        energies. However, with major differences between the electronic and vibrational structure of molecules, the question of how the resource requirements
        of computing anharmonic, vibrational states can be reduced remains relatively unexplored compared to
        its electronic counterpart. Importantly, bosonic commutation relations, distinguishable Hilbert spaces
        and vibrational coordinates allow manipulations of the vibrational system that can be exploited to minimize resource requirements.
        In this work, we investigate the impact of different coordinate systems
        and measurement schemes on the number of measurements needed to estimate
        anharmonic, vibrational states for a variety of three-mode
        (six-mode) molecules. We demonstrate an average of 3-fold (1.5-fold), with up to 7-fold (2.5-fold),
        reduction in the number of measurements required by employing appropriate coordinate transformations,
based on an automized construction of qubit Hamiltonians from a conventional vibrational structure program.
    \end{abstract}

    \maketitle

    \section{Introduction}\label{sec:introduction}

    The simulation of many-body quantum systems on quantum computers is a promising candidate to achieve computational
    advantages in both academic and practical applications \cite{daley_practical_2022,lee_is_2022,elfving_how_2020, aspuru-guzik_simulated_2005}.
    However, current quantum devices are subject to noise and errors to a degree that restricts the available computational resources.
    While practical quantum advantages are yet to be demonstrated, hybrid quantum-classical algorithms such as the
    \ac{vqe} are expected to provide such demonstrations ~\cite{anand_quantum_2022,babbush_exponentially_2016,babbush_exponentially_2017,mcclean_theory_2016,noauthor_variational_nodate}.
    The electronic structure problem on quantum computers has in recent years been investigated greatly while the
    vibrational structure problem remains relatively unexplored~\cite{sawaya_near-_2021,sawaya_resource-efficient_2020,mcardle_digital_2019,ollitrault_hardware_2020}.
    Importantly, it has been suggested that classically intractable vibrational structure problems may be solved on
    quantum computers prior to their electronic equivalents~\cite{sawaya_near-_2021}. Estimating eigenstates of many-body
    Hamiltonians using the \ac{vqe} requires the sampling of expectation values of Pauli operators with a trial state.
    However, it has been shown empirically
    that the estimation of the electronic ground state of small molecules to chemical accuracy requires infeasible
    runtimes due to sampling overhead, which is called the measurement problem~\cite{gonthier_identifying_2020,wang_minimizing_2021}.
    Different measurement schemes have been employed to reduce the sampling overhead, including the grouping of Pauli products
    into mutually commuting sets~\cite{zhao_measurement_2020,verteletskyi_measurement_2020,yen_deterministic_2022,bansingh_fidelity_2022,yen_measuring_2020, hamamura_efficient_2020,crawford_efficient_2021,choi_improving_2022,izmaylov_revising_2019,izmaylov_unitary_2019},
    classical shadow tomography \cite{hadfield_measurements_2020,huang_efficient_2021,hadfield_adaptive_2021,wu_overlapped_2021},
    low-rank factorizations of the electronic Hamiltonian \cite{huggins_efficient_2021}
    and algebraic approaches \cite{yen_cartan_2021}. Current studies have
    however only investigated electronic
    Hamiltonians.
    The impact of measurement schemes on the measurement problem therefore remains relatively unexplored in the context of vibrational structure~\cite{gonthier_identifying_2020}.
    In contrast to the two-body electronic Hamiltonian, the many-body vibrational Hamiltonian contains up to $N$-body
    coupling terms between the vibrational modes and depends on the choice of vibrational coordinates.
    The choice of coordinate system impacts the vibrational interactions and may be exploited to obtain approximate mode decoupling.
    Such a decoupling affects the distribution of small and large terms,
    which in favorable cases decreases the importance of high-mode coupling terms.
    This can be seen as better decoupling of coordinates and fewer very important terms, but as all terms are effected,
    this does not necessarily imply optimal variances. In this paper, we investigate the impact of different measurement schemes
    and coordinate systems on the number of measurements needed to estimate the ground state
    of a variety of three- and six-mode molecules.
    The runtime of a ground state calculation depends on the number of measurements required to
    estimate the ground state energy, which is determined by the variance of the Hamiltonian.
    We therefore study the required number of measurements and
    demonstrate an average of 3-fold (1.5-fold), with up to 7-fold (2.5-fold), reduction in this number using appropriate coordinate
    transformations. These results are obtained using state of the art vibrational structure \acp{pes} and anharmonic wave function computations and an automized construction of qubit Hamiltonians from a conventional vibrational structure program.

    \section{Background}

    \subsection{Sampling expectation values}
    Consider a qubit Hamiltonian $H = \sum_{i} h_{i} P_{i}$,
    where $h_{i}$ denotes matrix elements and
    \begin{equation}
        P_{i} = \bigotimes_{j} \sigma_{j}^{\alpha}, \quad \alpha\in\{x,y,z\},
    \end{equation}
    denotes a product of Pauli operators.
    In the \ac{vqe}, the ground state energy functional is estimated as the expectation value of the Hamiltonian with a
    trial state
    $\ket{\psi(\boldsymbol{\theta})}$,
    \begin{equation}
        E(\boldsymbol{\theta}) = \bra{\psi(\boldsymbol{\theta})} H \ket{\psi(\boldsymbol{\theta})} = \sum_{i} h_{i} \bra{\psi(\boldsymbol{\theta})} P_{i} \ket{\psi(\boldsymbol{\theta})}. \label{eq:equation2}
    \end{equation}
    To reduce the number of measurements, the Hamiltonian may be grouped into mutually commuting sets
    \begin{equation}
        S = \big\{ H_{\alpha} = \sum_{\beta} H_{\alpha}^{\beta} \;\big|\; [H_{\alpha}^{\beta}, H_{\alpha}^{\gamma}] = 0 \big\}
    \end{equation}
    such that
    $H = \sum_{\alpha} H_{\alpha}$. Each set, $H_{\alpha}$, may be rotated into a diagonal basis and 
    the operators of the set measured simultaneously.
    Let $m_{\alpha}$ denote the number of measurements required to estimate the energy contribution from $H_\alpha$
    and $\mathcal{M} = \sum_{\alpha} m_{\alpha}$ the total number of measurements.
    By optimizing $\{m_{\alpha}\}$ the number of measurements required for a given
    precision $\epsilon$ reads
    \begin{equation}
        \mathcal{M} = \left(\frac{\sum_{\alpha}\sqrt{\var(H_{\alpha})}}{\epsilon}\right)^2
        \label{eq:measurements}
    \end{equation}
    if $m_\alpha$ are optimally allocated for each group~\cite{yen_deterministic_2022}. With the total number of
    measurements depending on the variance of each group, the measurements may be optimized by optimizing the total group variances.

    \subsection{Measurement schemes}\label{sec:measurement-schemes}
    Qubit-space measurement schemes group mutually commuting Pauli products into measurable sets according to
    either \ac{qwc} or \ac{fc}~\cite{verteletskyi_measurement_2020,izmaylov_revising_2019,jena_pauli_2019,zhao_measurement_2020,yen_measuring_2020,hamamura_efficient_2020,yen_deterministic_2022}.
    In the \ac{qwc} scheme, the Pauli products commute locally for each qubit subspace, whereas the \ac{fc} scheme
    only requires commutativity of the full tensor products.
    It is important to note that the commutativity scheme alone (i.e. \ac{qwc} or \ac{fc}) does not define an algorithm for grouping operator terms.
    For definiteness, we consider the \ac{si} approach~\cite{crawford_efficient_2021,yen_deterministic_2022} in combination with \ac{qwc} or \ac{fc} 
    and denote the resulting algorithms as \acs{siqwc} and \acs{sifc}, respectively.
    In the \ac{si} approach one starts by ordering the operator terms (Pauli strings) in decreasing order according to the absolute value of their coefficients. The term
    with the largest coefficient is added to the first group. The algorithm then iterates through the remaining terms, which are added to the first group if they commute
    (in the \ac{qwc} or \ac{fc} sense) with all other terms in that group. When all terms have been checked, the algorithm starts over by generating a second group and so on
    until no terms are left. Despite its simplicity, the \ac{si} algorithm has been demonstrated to outperform
    recent classical shadow tomographic methods~\cite{yen_deterministic_2022}. Thus, the \ac{si} algorithm will be used in
    this work.

    The qubit-wise commutativity of two terms implies that the terms commute fully, although the converse is not true. In that sense, \ac{qwc} can be said to form
    a subset of \ac{fc}. It should however be kept in mind that the algorithms \acs{siqwc} and \acs{sifc} will generally produce different groupings of
    a given set of operator terms.

    After grouping the operator terms, each group is diagonalized. In the \ac{qwc} case, the diagonalization can be performed using only one-qubit gates, while
    the \ac{fc} case will generally require both one- and two-qubit gates.
    The \ac{qwc} scheme thus has a prima facie advantage
    in terms of circuit depth. For electronic Hamiltonians it has however been shown that this advantage
    is outweighed by other factors making the \ac{fc} scheme preferable~\cite{yen_deterministic_2022,bansingh_fidelity_2022} in that case.

    \section{Vibrational structure}

    \subsection{Vibrational Hamiltonian and potential energy surfaces}
    Consider a molecule with $M$ vibrational modes described by a set of mass-weighted
    \acp{nc} (or other orthogonal coordinates) $q_{m}$.
    Neglecting the Coriolis and pseudopotential terms of the Watson operator~\cite{watsonSimplificationMolecularVibrationrotation1968,watsonVibrationrotationHamiltonianLinear1970},
    the vibrational Hamiltonian reads
    \begin{equation}
        H = -\frac{1}{2}\sum_{m=1}^{M} \frac{\partial^2}{\partial q_{m}^2} + \sum_{\mathclap{m=1}}^{M} V(q_{m}) + \sum_{\mathclap{m<m'}}^{M}V (q_{m}, q_{m'}) + ...
    \end{equation}
    where the potential in principle contains terms coupling up to $M$ modes simultaneously. The \ac{pes},
    i.e. the vibrational potential, may be obtained using a variety of methods, e.g. using Taylor
    expansion or the \ac{adga}~\cite{sparta_adaptive_2009}. A Taylor expansion of the \acp{pes} around the equilibrium
    geometry provides the \acp{pes} in an economical form but suffers from the limited reliability of such expansions
    including limited radius of convergence and high risk of providing a variationally unbound potential. In contrast,
    the \ac{adga} is a robust and accurate black-box procedure, which builds the
    potential according to the need as determined from a vibrational calculation. The \ac{adga} uses very inexpensive vibrational self-consistent field densities
    to iteratively sample the surface in a physically motivated way.
    The \ac{pes} construction is thereby independent of the following correlated vibrational wave function computation, which is our main focus.
    After generating the \acp{pes}, the
    vibrational Hamiltonian may be transformed into a qubit Hamiltonian using existing encoding algorithms~\cite{mcardle_digital_2019,sawaya_resource-efficient_2020,ollitrault_hardware_2020}.

    \subsubsection{Vibrational coordinates}\label{subsec:vibrational-coordinates}
    A standard choice of coordinates is \acp{nc} which provide excellent first order insights into the
    vibrations of stiff molecules close to their equilibrium structures. The benefit of \acp{nc}
    for stiff molecules close to their equilibrium structures
    is that they decouple the Hamiltonian to second order in distortions from the equilibrium structure.
    However, the use of \acp{nc} does not suppress higher-order terms
    (in fact, the contrary might be true away from equilibrium).
    This motivates the search for other types of coordinates that ensure some degree of decoupling of the Hamiltonian.
    Such coordinates include various kinds of curvilinear coordinates (e.g. bond angles and lengths or polyspherical coordinates),
    which are often deemed chemically relevant.
    These coordinates typically reduce the couplings in the potential energy at
    the price of increasing the coupling level and overall complexity of the kinetic energy. To avoid this latter complication, we focus
    on rectilinear coordinates that are derived from \acp{nc} by orthogonal transformations. 
    All such coordinates are uniquely defined by a rectangular matrix $\mathbf{Q}$ with orthogonal columns that contains expansion 
    coefficients in terms of mass weighted Cartesian displacements (see \cref{app:coordinate_details}).
    They include optimized coordinates \cite{yagi_optimized_2012,thomsen_optimized_2014}
    obtained by minimizing the \ac{vscf} energy for a given \ac{pes} and localized coordinates \cite{jacob_localizing_2009} obtained by applying some
    localization scheme to a set of \acp{nc}.
    A pragmatic and inexpensive compromise is offered by the so-called \acp{holc} \cite{klinting_hybrid_2015}, which minimize a cost function including
    an energy term and a localization term:
    \begin{align}
        \mathbf{Q}_\textsc{holc} = \argmin_{\mathbf{Q}} \big( w_{\textsc{e}} E(\mathbf{Q}) + w_{\textsc{l}} L(\mathbf{Q}) \big).
    \end{align}
    Here, $\mathbf{Q}$ is varied over matrices that are related to the 
    normal coordinate matrix $\mathbf{Q}_\textsc{nc}$ by orthogonal transformations.
    The energy term $E(\mathbf{Q})$ is taken to be the \ac{vscf} energy for a second-order
    Taylor approximation of the \ac{pes}. This energy depends on the
    choice of coordinates, as indicated. The localization term (penalty term) $L(\mathbf{Q})$
    can be defined in various ways (we use the simple localization scheme
    described in Ref. \citenum{klinting_hybrid_2015}; see \cref{app:coordinate_details} for details).
    
    We note that setting
    $w_{\textsc{l}} = 0$ yields \acp{nc}, while
    a large value of $w_{\textsc{l}}$ yields very localized
    (but not necessarily meaningful) coordinates.
    We emphasize the fact that, as was mentioned in the introduction, approximate decoupling of the
    Hamiltonian does not necessarily provide optimal variances.

    \subsubsection{Distinguishability of vibrational modes}\label{subsec:vibrational-heuristics}
    Since vibrational modes are distinguishable (in constrast to electrons), the vibrational Hilbert space
    factorizes into distinguishable one-mode subspaces. This factorization allows for further grouping schemes
    that are not accesible to electronic Hamiltonians. To see this, we write the vibrational Hamiltonian as
    \begin{equation}
        H = \sum_{\mathclap{\mathfrak{m}\in\mathfrak{M}}} H^{\mathfrak{m}}, \quad H^{\mathfrak{m}} = \sum_{i} \tilde{H}^{\mathfrak{m}}_i.
    \end{equation}
    The sum runs over sets of modes ($\mathfrak{m}$), which we denote as \acp{mc}, and the set of all \acp{mc} ($\mathfrak{M}$) is 
    referred to as the \ac{mcr}. The operator $H^{\mathfrak{m}}$ is the sum of all terms that act non-trivially on the modes contained in $\mathfrak{m}$.
    Consider, for example, the operators $\tilde{H}^{(0,1)} = A_{0} \otimes A_{1} \otimes \mathds{1} \otimes \mathds{1} \otimes \mathds{1}$
    and
    $\tilde{H}^{(2,3,4)} = \mathds{1} \otimes \mathds{1} \otimes A_{2} \otimes A_{3} \otimes A_{4}$. Since the \acp{mc} $(0,1)$ and $(2,3,4)$ are
    disjoint (non-overlapping), the operators commute trivially. However, operators within the same \ac{mc} do not generally commute.
    We thus propose the following grouping scheme: First, non-overlapping \acp{mc} are combined into larger batches of terms. Then,
    each such batch is grouped in a trivially parallel fashion using either \ac{siqwc} or \ac{sifc}.
    The resulting algorithms are denoted as SI/QWC/MCR and SI/FC/MCR,
    respectively. Although the \ac{mc} logic does not solve the grouping problem on its own, it provides
    a sensible scheme that cheaply divide the problem into batches of subproblems
    (see \cref{appendix:mcr_grouping_five_mode_systems_example} for an example).

    Note that the \ac{mc} based commutativity is included in \ac{qwc} and \ac{fc} but that the concrete algorithms (SI/QWC/MCR and SI/FC/MCR)
    will generally result in different groups compared to SI/QWC and SI/FC.

    \section{Computational details}

    A total of 18 molecules were considered, including nine triatomic (three-mode) systems and nine tetratomic (six-mode) systems.
    All electronic structure calculations were performed at the CCSD(F12*)(T)/cc-pVDZ-F12 level~\cite{hattigCommunicationsAccurateEfficient2010,petersonSystematicallyConvergentBasis2008} of theory
    as implemented in the Turbomole~\cite{TurbomoleV7} program suite.
    The geometry of each molecule was optimized using numerical gradients, after which a numerical Hessian was
    computed and used to generate \acp{nc} and \acp{holc} with a series of localization parameters, $w_\textsc{l}$.
    Coordinate generation was performed using the MidasCpp program \cite{artiukhinMidasCpp2022}, which was also used
    to construct electronic PESs with the ADGA algorithm for each coordinate set
    as well as \ac{vscf} and \ac{fvci} computations.
    Following the \ac{pes} construction,
    \ac{vscf} calculations were carried out in large B-spline bases. 
    The resulting \ac{vscf} modals were then used as a basis in conventional \ac{fvci} calculations. We studied the
    convergence of the \ac{fvci} energy in terms of the number of \ac{vscf} modals and found that four (three) modals recovered
    a large fraction of the \ac{fvci} correlation energy for the three-mode (six-mode) systems (see \cref{appendix:convergence} for details).
    These bases were therefore chosen as a good balance between accuracy and CPU-time for the variance computations,
    which is the computational bottleneck of our locally developed Python3 code for this purpose.
    Having determined appropriate basis sizes, the Hamiltonians were represented in a suitable format, a process that 
    is fully automatized within MidasCpp (see \cref{app:operator_format} for a few considerations in this regard).
    The Hamiltonian and the \ac{fvci} wave functions were finally transformed to
    a qubit representation and the variance computed using the aforementioned variance code. A direct encoding was
    used due to its simplicity.
    We note that several encoding methods other than the direct mapping
    have been studied. In particular, compact encodings allow a reduction in the number of qubits which, depending on the
    problem, either decrease or increase circuit depths \cite{sawaya_resource-efficient_2020}.
    In this preliminary study, however, we focus on the direct mapping.
    For each molecule and coordinate system, the measurement groups were generated using the \acs{siqwc}, \acs{sifc},
    \ac{si}/\ac{qwc}/\ac{mcr} and \ac{si}/\ac{fc}/\ac{mcr} algorithms as described in \cref{sec:measurement-schemes,subsec:vibrational-heuristics} respectively.

    \begin{figure}
        \includegraphics[width=\columnwidth]{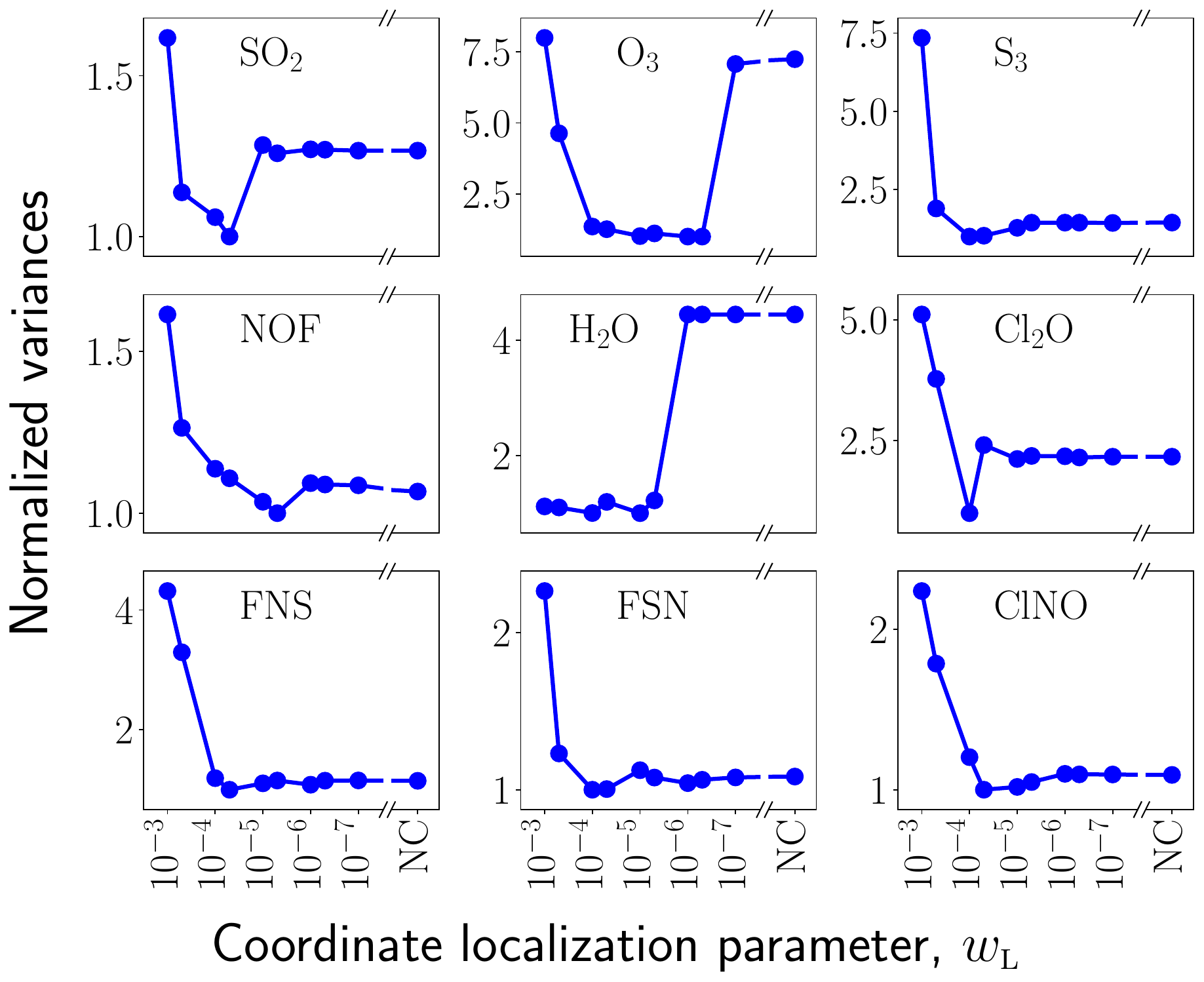}
        \caption{Variances of the three-mode molecules normalized relative to the minimum variance
        for all coordinate systems. For each coordinate system, the variance depicted is the optimal variance
        considering both the \ac{siqwc} and \ac{sifc} algorithms.
        Each \ac{holc} coordinate system is represented by its localization parameter $w_{\textsc{l}}$ and \ac{nc}
        refers to \aclp{nc} (equivalent to $w_{\textsc{l}} = 0$).
        Note the broken first axis that places the \ac{nc} data points alongside the \ac{holc} data.}
        \label{fig:variances_3}
    \end{figure}

    \begin{figure}
        \includegraphics[width=\columnwidth]{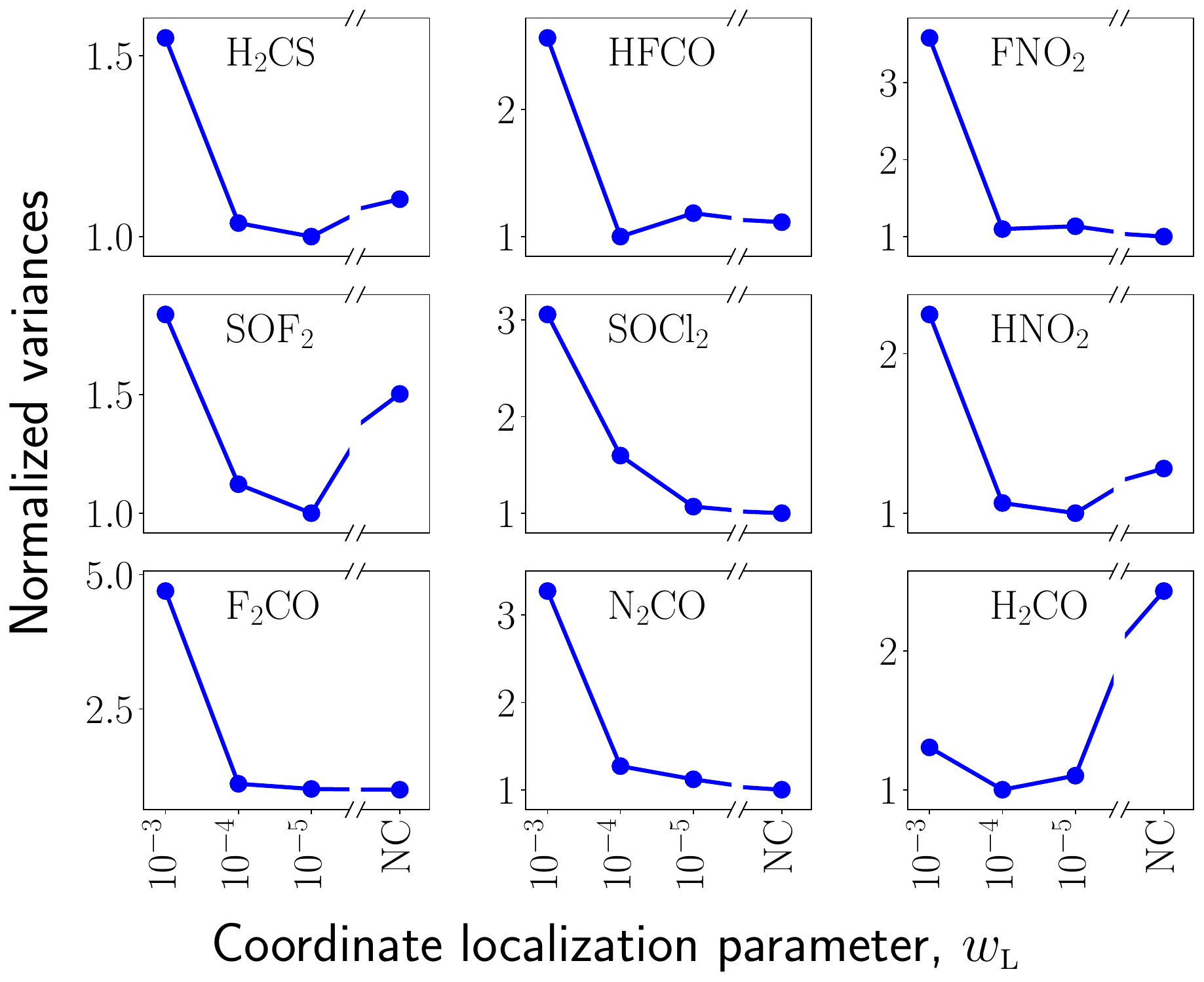}
        \caption{Variances of the three-mode molecules normalized relative to the minimum variance
        for all coordinate systems. For each coordinate system, the variance depicted is the optimal variance
        considering both the \ac{siqwc} and \ac{sifc} algorithm.
        Each \ac{holc} coordinate system is represented by its localization parameter $w_{\textsc{l}}$ and \ac{nc}
        refers to \aclp{nc} (equivalent to $w_{\textsc{l}} = 0$).
        Note the broken first axis that places the \ac{nc} data points alongside the \ac{holc} data.}
        \label{fig:variances_6}
    \end{figure}

    \section{Results} \label{sec:results}

        We initially consider the \ac{siqwc} and \ac{sifc} algorithms for each molecule and each set of coordinates.
        Selecting the smaller of the two variances (\ac{siqwc} or \ac{sifc}), \cref{fig:variances_3,fig:variances_6} are obtained.
        Almost all molecules exhibit the largest variances for strongly localized coordinates with localization parameter
        $w_{\textsc{l}} = \num{1.0e-3}$. Such behaviour is not surprising,
        since strongly localized coordinates are not necessarily physically meaningful.
        Localization of coordinates yields a smaller variance for some of the molecules but not all,
        and as such no choice of coordinates is systematically better than the others.

        \begin{figure}
            \includegraphics[width=\columnwidth]{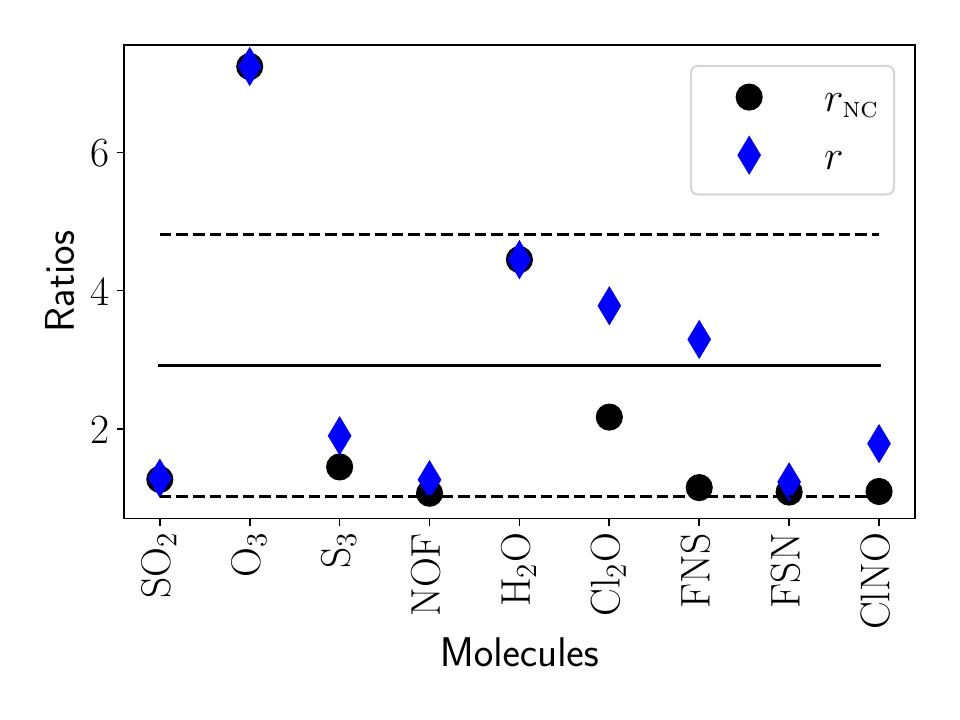}
            \caption{The ratio for the three-mode molecules between the variances of the worst and best
                choice of coordinates [\cref{eq:red_max}] in blue diamonds,
                along with the ratio between the variances of the optimal choice of coordinates and \acp{nc} [\cref{eq:red_nc}] in black dots.
                All coordinate systems are compared excluding the localization parameter
                $w_{\textsc{l}} = \SI{1.0e-3}{}$ since these coordinates were inferior for all molecules.}
            \label{fig:reductions_3}
        \end{figure}
    
        \begin{figure}
            \includegraphics[width=\columnwidth]{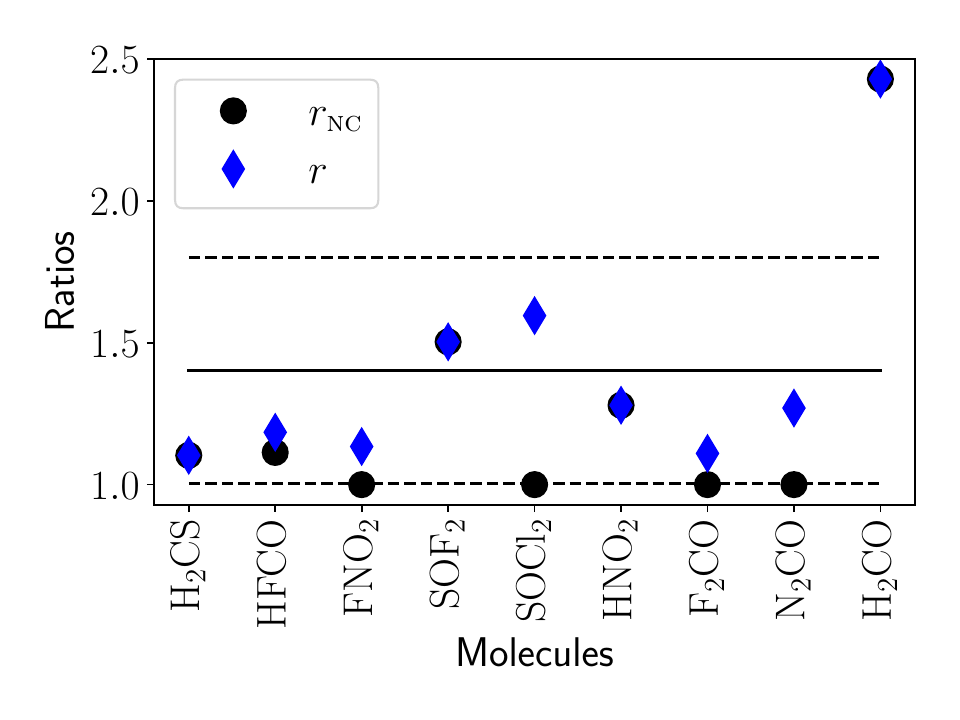}
            \caption{The ratio for the six-mode molecules between the variances of the worst and best
                choice of coordinates [\cref{eq:red_max}] in blue diamonds,
                along with the ratio between the variances of the optimal choice of coordinates and \acp{nc} [\cref{eq:red_nc}] in black dots.
                All coordinate systems are compared excluding the localization parameter
                $w_{\textsc{l}} = \SI{1.0e-3}{}$ since these coordinates were inferior for all molecules.}
            \label{fig:reductions_6}
        \end{figure}

        \subsection{Invariance and reductions}
        To further estimate the impact of coordinate transformations, two ratios are studied.
        Using the data displayed in \cref{fig:variances_3,fig:variances_6} the lowest and highest variance is calculated
        for each molecule with $w_{\textsc{l}} = \num{1.0e-3}$ excluded
        as it is vastly inferior for most of the molecules.
        The first ratio compares the lowest and highest of these variances,
        \begin{subequations}
        \begin{align}
            r &= \frac{\textup{maximum variance}}{\textup{minimum variance}}, \label{eq:red_max}
            \intertext{while the second ratio compares the variance of \acp{nc} to the minimum variance,}
            r_{\textsc{nc}} &= \frac{\textup{NC variance}}{\textup{minimum variance}}. \label{eq:red_nc}
        \end{align}
        \end{subequations}
        That is for \ce{NOF} the ratio $r$ is that of the variances at $w_\textsc{l} = \num{5e-3}$ and
        $w_\textsc{l} = \num{5e-5}$, as these are the maximum and minimum variances for the molecule
        when excluding the consistently unfavourable $w_\textsc{l} = \num{1e-3}$ data point.
        Likewise $r_\textsc{nc}$ is variance ratio of \acp{nc} to $w_\textsc{l} = \num{5e-5}$.

        If $r_\textsc{nc} > 1$, \acp{nc} provide a larger variance compared to HOLCs and thus HOLCs are preferential.
        In contrast, if $r_\textsc{nc} = 1$, \acp{nc} provide a smaller variance compared to HOLCs and thus \acp{nc} are preferential.
        $r$ therefore serves as a measure of the overall impact on variance of coordinate transformation,
        whereas $r_\textsc{nc}$ measures the extent to which \acp{holc} improve the variance compared to \acp{nc}.
        These reductions measures are depicted in \cref{fig:reductions_3,fig:reductions_6}.

        The mean reduction for the three-mode (six-mode) molecules is around 3 (1.5) with relatively large standard deviations.
        The largest improvements for the three-mode (six-mode) molecules amounts to around a 7-fold (2.5-fold) reduction.
        Meanwhile, some of the molecules exhibit approximate invariance under reasonable coordinate transformations
        ($r$ close to $1$).

        \subsection{Molecular symmetry}
\begin{figure*}
        \includegraphics[width=.9\textwidth]{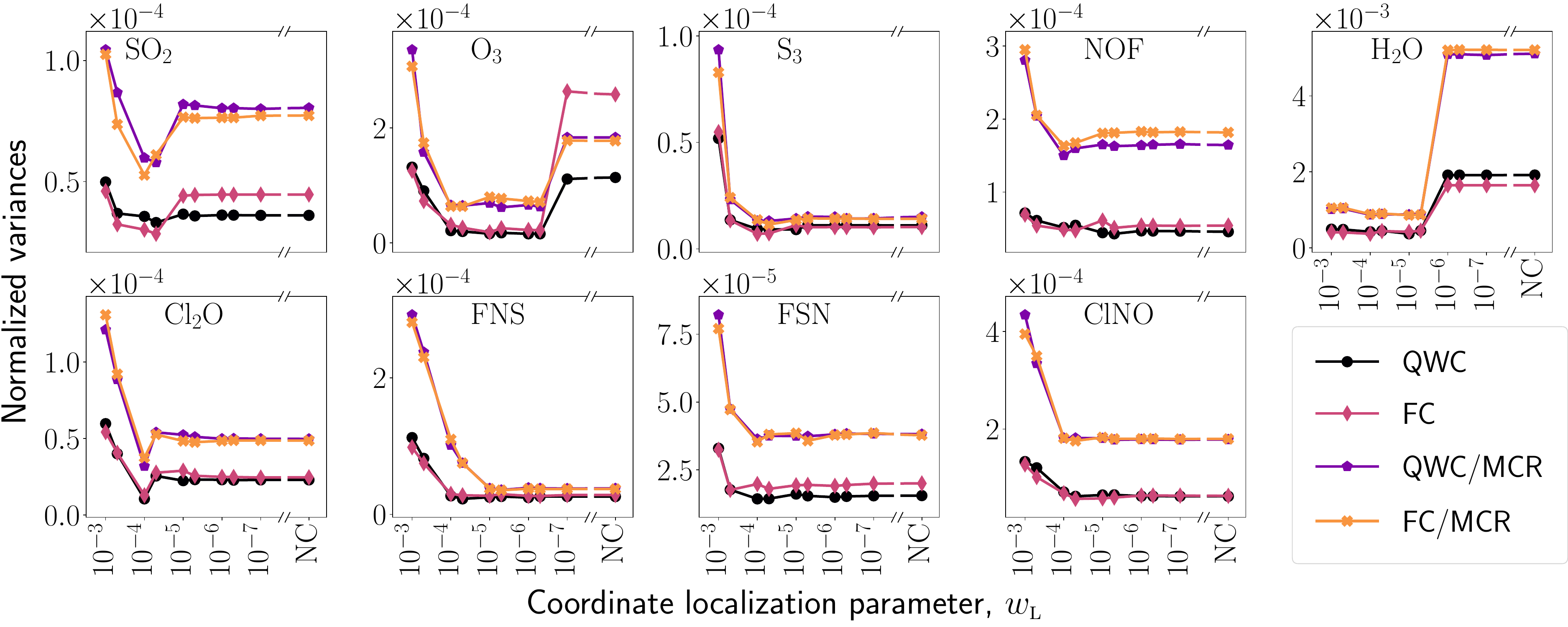}
        \caption{Variances of the three-mode molecules for each grouping scheme (\cref{sec:measurement-schemes,subsec:vibrational-heuristics}).
        The \ac{holc} coordinate systems are defined by their localization parameter $w_{\textsc{l}}$,
        with \ac{nc} denoting \aclp{nc} (equivalent to $w_{\textsc{l}} = 0$).
        Note the broken first axis that places the \ac{nc} data points alongside the \ac{holc} data.
        All grouping schemes employ \acl{si} with the commutator schemes indicated in the legend.
        In general, using \acl{si} for all terms (not using MCR grouping) produces the smallest variances and the
            discrepancy between commutativity schemes is small.}
        \label{fig:groups_3}
    \end{figure*}

        We see no clear pattern relating to point group symmetry
        as exemplified by the molecules \ce{O3} and \ce{S3}. Despite being extremely similar with respect
        to symmetry and structure, they behave quite differently with respect to the computed variances
        (see \cref{fig:reductions_3,fig:variances_3,fig:groups_3}).
        These results highlight the complexity of the interdependence between
        the choice of coordinates, the details of the \ac{pes} and the vibrational structure of the molecule.

    \begin{figure*}
        \includegraphics[width=.9\textwidth]{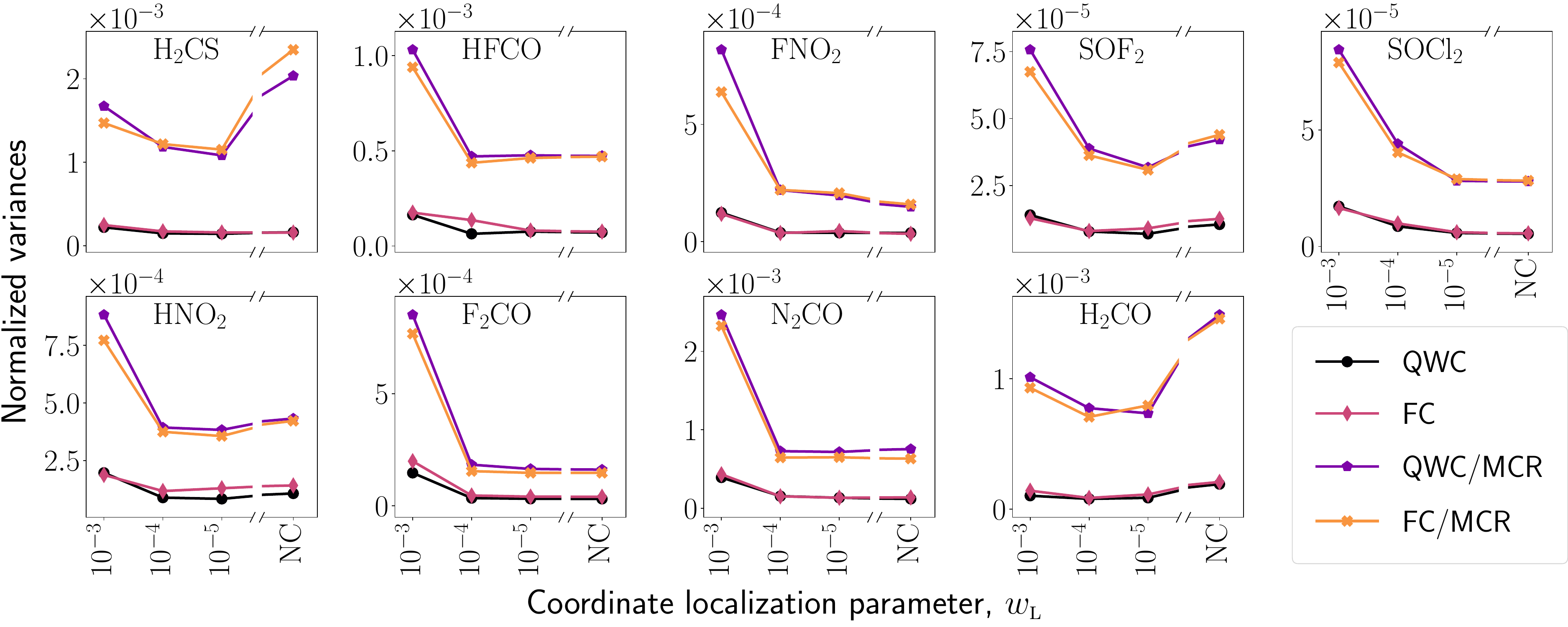}
        \caption{Variances of the six-mode molecules for each grouping scheme (\cref{sec:measurement-schemes,subsec:vibrational-heuristics}).
        The \ac{holc} coordinate systems are defined by their localization parameter $w_{\textsc{l}}$,
        with \ac{nc} denoting \aclp{nc} (equivalent to $w_{\textsc{l}} = 0$).
        Note the broken first axis that places the \ac{nc} data points alongside the \ac{holc} data.
        All grouping schemes employ \acl{si} with the commutator schemes indicated in the legend.
        In general, using \acl{si} for all terms (not using MCR grouping) produces the smallest variances and the
            discrepancy between commutativity schemes is small.}
        \label{fig:groups_6}
    \end{figure*}

    \subsection{\ac{siqwc} and \ac{sifc} schemes}
    To study the \ac{siqwc} and \ac{sifc} schemes, the variances were calculated for each measurement scheme which is depicted
    in \cref{fig:groups_3,fig:groups_6}. In contrast to a corresponding electronic structure study \cite{yen_deterministic_2022},
    no commutativity scheme systematically outperforms the other across molecules and coordinates.
    In particular, no appreciable return on investment is observed for the additional circuit depth of \ac{fc}
    compared to \ac{qwc}.
    The vibrational operators are very different from their electronic counterparts
    with the electronic Hilbert space spanned by spin-orbitals and the vibrational
    Hilbert space of modal basis functions.
    This might explain the difference in relative performance of the \ac{si}/\ac{qwc} and \ac{si}/\ac{fc} algorithms
    when comparing vibrational and electronic structure.

        In \cref{fig:groups_3} there are some significant differences in the performance of the two commutativity schemes,
        e.g. \ce{SO2} and \ce{O3}, is observed for the three-mode systems.
        This is however not observed for the six-mode systems (\cref{fig:groups_3}),
        which might be due to a peculiarity of the three-mode Hilbert space.
        With three modes there is a single \ac{mc} in the \ac{mcr} that contains a vast majority of all terms in the Hamiltonian.
        For the six-mode systems the number of terms are more evenly distributed among the \acp{mc},
        which may explain the different observed behaviours of the commutativity schemes
        as operators from two non-overlapping \acp{mc} commute (see \cref{subsec:vibrational-heuristics}).

    \subsection{Vibrational heuristics for sorting} \label{sec:heuristics}
    From \cref{fig:groups_3,fig:groups_6} it is clear that \ac{siqwc} and \ac{sifc} outperform \acs{siqwcmcr} and \acs{sifcmcr}.
    \Ac{siqwc} and \ac{sifc} thus seem to be the best options,
    which in some sense is not surprising. The \ac{mcr} methods essentially
    divide the set of all Hamiltonian terms into smaller subsets, in a chemically motivated fashion.
    The \ac{si} algorithm then sorts these subsets in order to obtain the final groups, hence the sorting is restricted,
    meaning that the sorting has less flexibility.
    For larger molecules this might however be an advantage if the sorting overhead becomes significant due to the non-linear
    scaling of sorting algorithms.
    This scaling implies that in general sorting of multiple small sets is faster than the sorting of one large set.
    Additionally, the sorting of smaller sets is trivially parallelizable in contrast to a single sorting of the full set.
    As seen for \ce{S3} and \ce{FNS} the \ac{mcr} methods has the potential to yield groupings of comparable variance to plain \ac{si}
    hence it is possible that the \ac{mcr} methods might yield a faster overall time to solution for large molecules.
    For the three-mode systems the vast majority of terms are contained within the \ac{mc} $\mathfrak{m} = (1,2,3)$,
    hence the potential speed-up in group generation is negligible in this case.
    However, for larger molecules, the number of combinations of higher-order terms increase rapidly.
    
    \begin{figure}
           \includegraphics[width=\columnwidth]{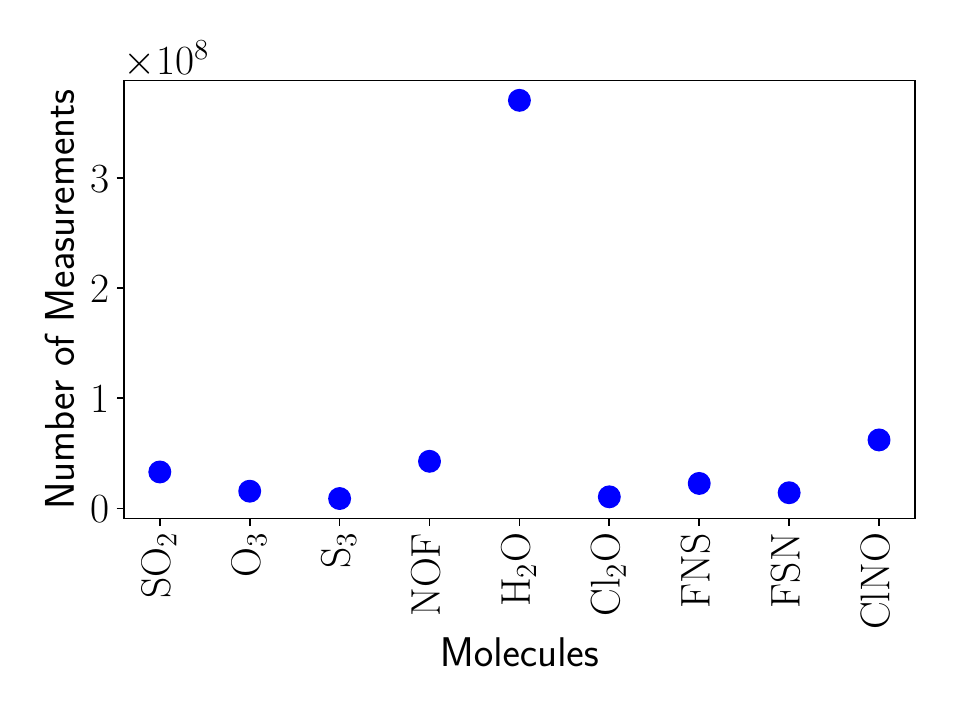}
        \caption{The number of measurements for estimating a single energy evaluation for each of the three-mode benchmark molecules
            using the \ac{si}/\ac{qwc} sorting scheme.
            The three-mode bases includes 4 modals per mode.}
        \label{fig:measurements_3}
    \end{figure}

    \begin{figure}
           \includegraphics[width=\columnwidth]{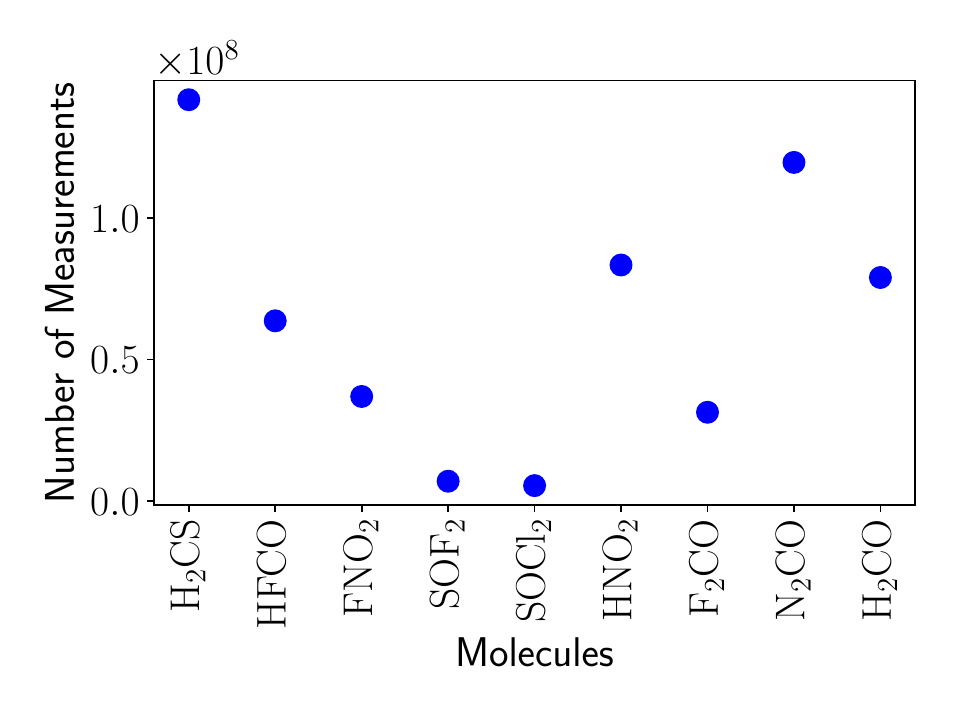}
        \caption{The number of measurements for estimating a single energy evaluation for each of the six-mode benchmark molecules
            using the \ac{si}/\ac{qwc} sorting scheme.
            The six-mode bases include 3 modals per mode.}
        \label{fig:measurements_6}
    \end{figure}

    \subsection{Measurement numbers}
    Using the \ac{si}/\ac{qwc} variances with the optimal choice of coordinates, for each molecule,
    the number of measurements for estimating a single energy evaluation are shown in \cref{fig:measurements_3,fig:measurements_6}.
    These measurement numbers are calculated using \cref{eq:measurements} with an error of $\epsilon = \SI{1}{\micro\hartree}$
    corresponding to relevant accuracy for vibrational structure.
    The number of measurements required for both the three-mode and six-mode molecules falls in the order of magnitude range \numrange{1e7}{1e8}.
    That the measurement numbers for the three- and six-mode systems are so close on average is somewhat surprising
    considering the size of Hilbert space or the number of operator terms in the Hamiltonian.
    The dimensions of the Hilbert space are
    \begin{align}
        D &= N^M = \begin{cases}
            \num{64}  \quad &\text{for} \quad M=3, N=4, \\
            \num{729} \quad &\text{for} \quad M=6, N=3,
        \end{cases}
        \intertext{whereas the number of second quantized terms (in the format of \cref{app:operator_format})
        for a three-mode coupled Hamiltonian is}
        N_{\text{terms}} &=
        \binom{M}{3} N^6 + \binom{M}{2} N^4 + \binom{M}{1} N^2 \nonumber\\
        &= \begin{cases}
               \num{4912}  \quad &\text{for} \quad M=3, N=4, \\
               \num{15849} \quad &\text{for} \quad M=6, N=3.
        \end{cases}
    \end{align}

    It should, however, be recalled that the three-mode systems are dominated by a single three-mode coupling with many terms
    while the six-mode systems have a comparatively large number of mode couplings that each contribute a relatively small number of terms.
    This impacts the grouping of terms after the qubit transformation and, in turn, the variances and measurement numbers.
    The general structures of the two Hilbert spaces are thus quite different,
    and a naive argument based on the size of the Hilbert space or the number of operator terms alone is not sufficient.

    \section{Conclusion}
    \acresetall
    In this work, we have presented different vibrational coordinates that provide reductions in measurements
    for quantum computations of anharmonic vibrational wave funtions using realistic \acp{pes}.
    It was shown that an average of 3-fold (1.5-fold),
    with up to 7-fold (2.5-fold), reductions may be achieved by appropriately transforming the vibrational coordinates
    for three-mode (six-mode) molecules. One must emphasize the size of the molecules which were used in the benchmarks.
    Localized coordinates are generally favourable for relatively large molecules for which delocalized interactions may
    be negligible. Thus, localized coordinates should be further investigated in the context of larger systems to investigate
    the performance for larger molecules.

    Contrary to what has been observed for electronic structure, we see no systematic reduction in variance
    using the \ac{sifc} scheme compared to the \ac{siqwc} scheme.
    Such differences between electronic and vibrational structure may
    arise due to the different commutation relations (fermionic/bosonic). Additionally, the Hilbert spaces differ in
    structure, with the electronic Hilbert space spanned by spin-orbitals and the vibrational
    Hilbert space spanned by modal basis functions. Combining the \ac{si}
    algorithm with the grouping of non-overlapping interaction terms in the \ac{mcr} algorithm (\cref{subsec:vibrational-heuristics})
    does not provide better groupings compared to the plain \ac{si} algorithm.
    They are however still interesting because the sorting of multiple sets is trivially parallelizable
    which might be useful if the sorting overhead becomes significant, e.g. for large molecules.

    These results agree with previous conclusions for electronic structure where minimizing the
    total number of groups does not necessarily provide optimal variances.

    \section{Acknowledgements}
    O.C. acknowledges support from the Independent Research Fund Denmark through grant number 1026-00122B.
    The authors acknowledge funding from the Novo Nordisk Foundation through grant NNF20OC0065479.
    \Ac{pes} calculations were performed at the Centre for Scientific Computing Aarhus (CSCAA).

    \section{Data availability}
    The code related to the study is available upon reasonable request to the authors.

    \appendix

    \section{Details on coordinates}  \label{app:coordinate_details}
    For an $N$-atomic molecule we introduce a $3N$ dimensional column vector $\mathbf{d}$ containing the
    Cartesian displacements from the equilibrium geometry. We likewise introduce a $3N \times 3N$ diagonal matrix
    $\mathbf{G}$ of atomic masses,
    \begin{align}
        \mathbf{G} = \mathrm{diag}(M_{1}, M_{1}, M_{1}, \ldots, M_{N}, M_{N}, M_{N}).
    \end{align}
    A set of mass weighted displacements is then defined as $\mathbf{d}' = \mathbf{G}^{1/2} \mathbf{d}$
    and forms the basis for further coordinates $\mathbf{q}$ via an orthogonal transformation $\mathbf{Q}$:
    \begin{align}
        \mathbf{q} = \mathbf{Q}^{\mathsf{T}} \mathbf{d}'.
    \end{align}
    Normal coordinates correspond to the choice of $\mathbf{Q}$ that diagonalizes
    the mass weighted Hessian:
    \begin{align}
        (\mathbf{G}^{-1/2} \mathbf{F} \mathbf{G}^{-1/2}) \mathbf{Q} = \mathbf{Q} \mathbf{\Lambda},
    \end{align}
    where $\mathbf{\Lambda}$ is the vector of eigenvalues and $\mathbf{F}$ is the unweighted Hessian.
    Six eigenvectors (five for linear molecules) have zero eigenvalues and correspond to the overall 
    rotation and translation of the molecule. These eigenvectors are not included in $\mathbf{Q}$.
    New rectilinear coordinates $\mathbf{Q}'$ can be obtained by performing orthogonal transformations
    of the $\mathbf{Q}$ matrix:
    \begin{align}
        \mathbf{Q}' = \mathbf{Q} \mathbf{L}
    \end{align}
    Here, $\mathbf{L}$ is an $M \times M$ orthogonal matrix,
    where $M = (3N - 6)$ (non-linear molecules) or $M = (3N - 5)$ (linear molecules).
    In the \ac{holc} algorithm~\cite{klinting_hybrid_2015}, $\mathbf{L}$ is parametrized through Jacobi sweeps.
    The algorithm optimizes $\mathbf{L}$ such that the following cost function is minimized:
    \begin{equation}
        f(\mathbf{Q}') = w_{\textsc{e}} E(\mathbf{Q}') + w_{\textsc{l}} L(\mathbf{Q}').
    \end{equation}
    The energy term $E(\mathbf{Q}')$ is taken as the \ac{vscf} energy on a second-order approximate
    \ac{pes}, while we use an atomic localization term:
    \begin{align}
        L(\mathbf{Q}') &= - \sum_{k=1}^{M} \sum_{i=1}^{N} (C_{ik})^2 \\
        C_{ik} &= \sum_{\mathclap{\alpha = x,y,z}} (Q'_{i\alpha,k})^2 
    \end{align}

    \section{\ac{mcr} grouping for five-mode systems}
    \label{appendix:mcr_grouping_five_mode_systems_example}
    For a five-mode system with one-, two-, and three-mode couplings in the Hamiltonian, the grouping based on the
    \ac{mcr} may be obtained by grouping disjoint two- and three-body
    couplings along with one-body couplings. The two- and three-body couplings yield
    $\{H^{(i,j,k)}+H^{(m,n)}|(m,n)\not\subset (i,j,k)\}$
    while the one-body couplings yield $\{H^{(i)}\}$. An example is presented in the following:
    \begin{itemize}
        \item $S_0 = \{H^{(0,1,2)}, H^{(3,4)}\}$
        \item $S_1 = \{H^{(1,2,3)}, H^{(0,4)}\}$
        \item $S_2 = \{H^{(2,3,4)}, H^{(0,1)}\}$
        \item $S_3 = \{H^{(0,2,3)}, H^{(1,4)}\}$
        \item $S_4 = \{H^{(0,3,4)}, H^{(1,2)}\}$
        \item $S_5 = \{H^{(0,1,4)}, H^{(2,3)}\}$
        \item $S_6 = \{H^{(1,3,4)}, H^{(0,2)}\}$
        \item $S_7 = \{H^{(0,1,3)}, H^{(2,4)}\}$
        \item $S_8 = \{H^{(0,2,4)}, H^{(1,3)}\}$
        \item $S_9 = \{H^{(0)}, H^{(1)}, H^{(2)}, H^{(3)}, H^{(4)}\}$
    \end{itemize}
    The terms commute since their mode couplings are disjoint. The terms also exhibit maximal mode coupling since
    all modes are active in the terms. However, the operators for each mode coupling within an \ac{mcr}
    do not mutually commute and must be diagonalized. Using the \ac{si} algorithm, each \ac{mcr} group may be decomposed
    into subgroups which mutually commute. The \ac{mcr} grouping for the three- and six-mode molecules studied in this work
    are generated analogously to the five-mode system. Due to the more complex combinations for the six-mode system,
    however, we present the example of a five-mode system for simplicity.

    \section{Modal basis set convergence}
    \label{appendix:convergence}
\begin{figure*}
        \includegraphics[width=\columnwidth]{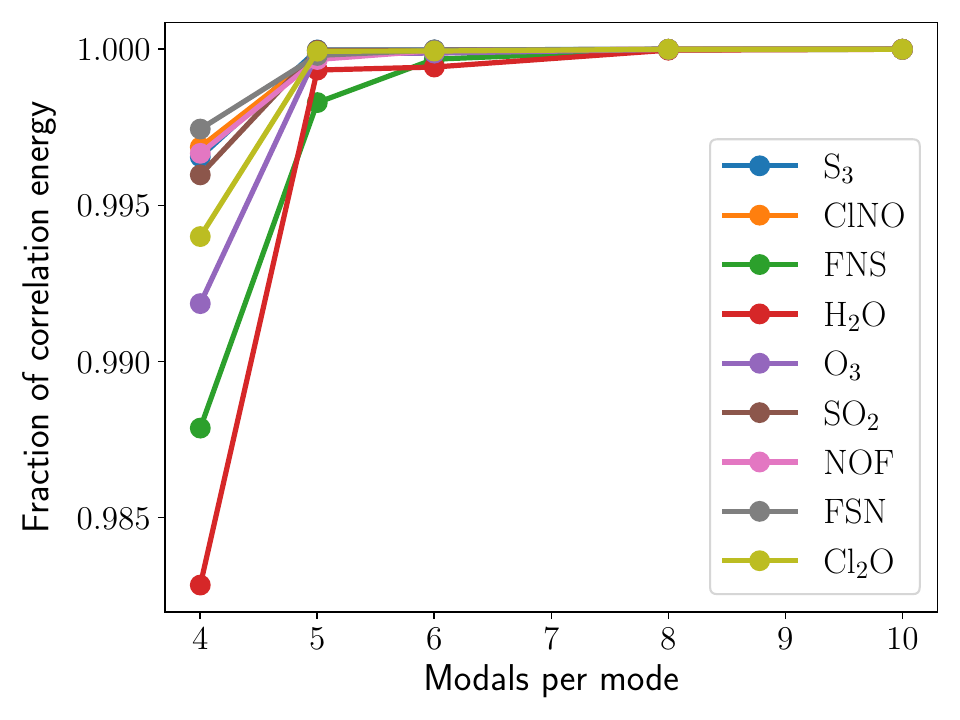}
        \caption{Convergence of the correlation energy with respect to the modal basis set size for the three-mode molecules.
        The displayed fraction is relative to the 10 modal computation for each molecule.}
        \label{fig:correlation_energy_3}
    \end{figure*}

\begin{figure*}
        \includegraphics[width=\columnwidth]{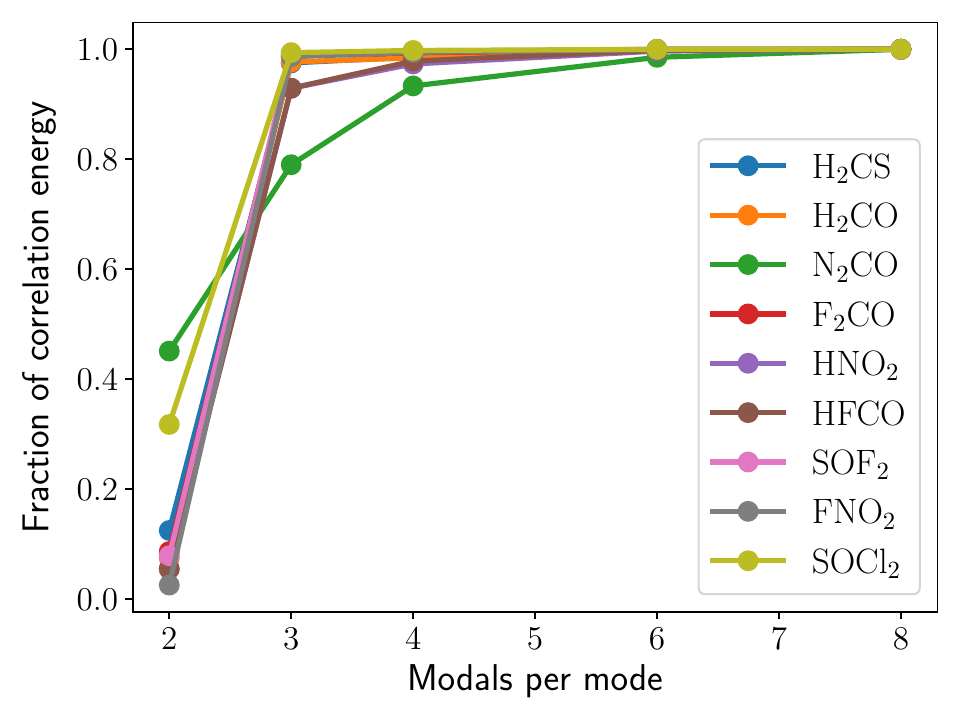}
        \caption{Convergence of the correlation energy with respect to the modal basis set size for the six-mode molecules.
            The displayed fraction is relative to the 8 modal computation for each molecule.}
        \label{fig:correlation_energy_6}
    \end{figure*}

    In order to test the convergence of the modal basis set dimensions, we calculate \ac{fvci} ground state correlation energies
    for up to ten (eight) modals per mode for the three-mode (six-mode) systems. The results are presented in Figs.
    \ref{fig:correlation_energy_3} and \ref{fig:correlation_energy_6}.
    As can be seen, a large fraction of the ground state correlation energies is recovered
    for four (three) modals per mode for the three-mode (six-mode) systems.


    \section{Operator format}  \label{app:operator_format}
    The Hamiltonian for the vibrational problem is usually represented in a \ac{sop} form,
    \begin{equation}
        \label{eq:oper_format}
        H = \sum_{t} c_t \prod_{m \in \mbf{m}_t} h^{m,t} = \sum_{\mbf{m}} H^{\mbf{m}},
    \end{equation}
    where $c_t$ is the coefficient for the term indexed by $t$. Any given term includes
    a product of one-mode operators $h^{m,t}$ for a set of modes $\mbf{m}_t$. The one-mode operators
    can be written as
    \begin{equation}
        h^{m,t} = \sum_{\mathclap{p^m q^m}} h^{m,t}_{p^m q^m} \crea{m}{p} \anni{m}{q}.
    \end{equation}
    The \ac{sop}
    format covers most practically relevant cases such as Taylor expanded potentials and more elaborate
    polynomial representations such as those generated by the \ac{adga} algorithm~\cite{sparta_adaptive_2009}.
    When designing algorithms
    for classical computers, it is extremely beneficial to keep the operator in the \ac{sop} form and never
    expand the products of one-mode operators. However, for the purpose of transforming the Hamiltonian
    into qubit format, we expand all product into simple strings of creating and annihilation operators.
    As an elementary example, consider the terms pertaining to a given two-mode combination,
    \begin{align}
        H^{mn} &= \sum_{t} c_t h^{m,t} h^{n,t} \nn
        &=
        \sum_{p^m q^m} \sum_{p^n q^n} \sum_{t} c_t \, h^{m,t}_{p^m q^m} h^{n,t}_{p^n q^n}
        \crea{m}{p} \anni{m}{q} \crea{n}{p} \anni{n}{q} \nn
        &\equiv \sum_{p^m q^m} \sum_{p^n q^n} H^{m,n}_{(p^m q^m)(p^n q^n)}
        \crea{m}{p} \anni{m}{q} \crea{n}{p} \anni{n}{q}. \label{eq:oper_expanded}
    \end{align}
    This example is readily generalized to terms containing more modes. A trivial but important
    point is that the number of coefficients in \cref{eq:oper_expanded} does not depend
    on the number of terms in the \ac{sop} expansion in \cref{eq:oper_format}.

    \bibliographystyle{ieeetr}
    \bibliography{../bib/vibqc.bib,../bib/madsgh.bib}

\end{document}

%% file: abbreviations.tex
\usepackage{acro}

\DeclareAcronym{mctdh}{
   short = MCTDH ,
   long = multiconfiguration time-dependent Hartree ,
}

\DeclareAcronym{nomctdh}{
   short = NOMCTDH ,
   long = non-orthogonal \ac{mctdh} ,
}

\DeclareAcronym{gmctdh}{
   short = G-MCTDH ,
   long = Gaussian-based \ac{mctdh} ,
}

\DeclareAcronym{mlgmctdh}{
   short = ML-GMCTDH ,
   long = multilayer Gaussian-based \ac{mctdh} ,
}

\DeclareAcronym{mlmctdh}{
   short = ML-MCTDH ,
   long = multilayer \ac{mctdh} ,
}

\DeclareAcronym{mpsmctdh}{
   short = MPS-MCTDH ,
   long = matrix product state \ac{mctdh} ,
}

\DeclareAcronym{vmcg}{
   short = vMCG ,
   long = variational multiconfiguration Gaussian ,
}

\DeclareAcronym{ms}{
   short = MS ,
   long = multiple spawning ,
}

\DeclareAcronym{ccs}{
   short = CCS ,
   long = coupled coherent states ,
}

\DeclareAcronym{mctdhn}{
   short = MCTDH[\textit{n}] ,
   long = systematically truncated multiconfiguration time-dependent Hartree ,
}

\DeclareAcronym{mrmctdhn}{
  short = MR-MCTDH[\textit{n}] ,
  long = multi-reference truncated multiconfiguration time-dependent Hartree ,
}

\DeclareAcronym{tdh}{
   short = TDH ,
   long = time-dependent Hartree ,
}

\DeclareAcronym{dmrg}{
   short = DMRG ,
   long = density matrix renormalization group,
}

\DeclareAcronym{tddmrg}{
   short = TD-DMRG ,
   long = time-dependent density matrix renormalization group,
}

\DeclareAcronym{scf}{
   short = SCF ,
   long = self-consistent field ,
}

\DeclareAcronym{casscf}{
   short = CASSCF ,
   long = complete active space self-consistent field ,
}

\DeclareAcronym{tdcasscf}{
   short = TD-CASSCF ,
   long = time-dependent \acl{casscf} ,
}

\DeclareAcronym{gasscf}{
   short = CASSCF ,
   long = generalized active space self-consistent field ,
}

\DeclareAcronym{tdgasscf}{
   short = TD-GASSCF ,
   long = time-dependent \acl{gasscf} ,
}

\DeclareAcronym{rasscf}{
   short = RASSCF ,
   long = restricted active space self-consistent field ,
}

\DeclareAcronym{tdrasscf}{
   short = TD-RASSCF ,
   long = time-dependent \acl{rasscf} ,
}

\DeclareAcronym{ormas}{
   short = ORMAS ,
   long = occupation-restricted multiple active space ,
}
\DeclareAcronym{tdormas}{
   short = TD-ORMAS ,
   long = time-dependent \acl{ormas} ,
}

\DeclareAcronym{mctdhf}{
   short = MCTDHF ,
   long = multiconfiguration time-dependent Hartree-Fock ,
}

\DeclareAcronym{occ}{
   short = OCC ,
   long = orbital-optimized coupled cluster ,
}

\DeclareAcronym{tdocc}{
   short = TD-OCC ,
   long = time-dependent \acl{occ} ,
}

\DeclareAcronym{nocc}{
   short = NOCC ,
   long = non-orthogonal orbital-optimized coupled cluster ,
}

\DeclareAcronym{oatdcc}{
   short = OATDCC ,
   long = orbital-adaptive time-dependent coupled cluster ,
}

\DeclareAcronym{fci}{
   short = FCI ,
   long = full configuration interaction ,
}

\DeclareAcronym{fvci}{
   short = FVCI ,
   long = full vibrational configuration interaction ,
}

\DeclareAcronym{cud}{
   short = CUD ,
   long = closed under de-exciation ,
}

\DeclareAcronym{fsmr}{
   short = FSMR ,
   long = full-space matrix representation ,
}

\DeclareAcronym{hh}{
   short = HH ,
   long = H\'enon-Heiles ,
}

\DeclareAcronym{ho}{
   short = HO ,
   long = harmonic oscillator ,
}

\DeclareAcronym{dop853}{
   short = DOP853 ,
   long = Dormand-Prince 8{(5,3)} ,
}

\DeclareAcronym{sm}{
   short = SM ,
   long = supplementary material ,
}

\DeclareAcronym{vscf}{
   short = VSCF ,
   long = vibrational self-consistent field ,
}

\DeclareAcronym{eom}{
   short = EOM ,
   long = equation of motion ,
   short-plural-form = EOMs ,
   long-plural-form = equations of motion ,
}

\DeclareAcronym{tdvp}{
   short = TDVP ,
   long = time-dependent variational principle
}

\DeclareAcronym{tdse}{
   short = TDSE ,
   long = time-dependent Schr{\"o}dinger equation ,
}

\DeclareAcronym{cc}{
   short = CC ,
   long = coupled cluster ,
}

\DeclareAcronym{vcc}{
   short = VCC ,
   long = vibrational coupled cluster ,
}

\DeclareAcronym{tdvcc}{
   short = TDVCC ,
   long = time-dependent vibrational coupled cluster ,
}

\DeclareAcronym{tdvci}{
   short = TDVCI ,
   long = time-dependent vibrational configuration interaction ,
}

\DeclareAcronym{vci}{
   short = VCI ,
   long = vibrational configuration interaction ,
}

\DeclareAcronym{ci}{
   short = CI ,
   long = configuration interaction ,
}

\DeclareAcronym{tdci}{
   short = CI ,
   long = time-dependent \acl{ci} ,
}

\DeclareAcronym{sq}{
   short = SQ ,
   long = second quantization ,
}

\DeclareAcronym{fq}{
   short = FQ ,
   long = first quantization ,
}

\DeclareAcronym{mc}{
   short = MC ,
   long = mode combination ,
}

\DeclareAcronym{mcr}{
   short = MCR ,
   long = mode combination range ,
   long-plural = s ,
}

\DeclareAcronym{pes}{
   short = PES ,
   long = potential energy surface
}

\DeclareAcronym{svd}{
   short = SVD ,
   long = singular value decomposition ,
}
\DeclareAcronym{adga}{
   short = ADGA ,
   long = adaptive density-guided approach ,
}

\DeclareAcronym{rhs}{
   short = RHS ,
   long = right-hand side ,
}

\DeclareAcronym{lhs}{
   short = LHS ,
   long = left-hand side ,
}

\DeclareAcronym{ivr}{
   short = IVR ,
   long = intramolecular vibrational energy redistribution ,
}

\DeclareAcronym{fft}{
   short = FFT ,
   long = fast Fourier transform ,
}

\DeclareAcronym{spf}{
   short = SPF ,
   long = single-particle function ,
}

\DeclareAcronym{lls}{
   short = LLS ,
   long = linear least squares ,
}

\DeclareAcronym{itnamo}{
   short = ItNaMo ,
   long = iterative natural modal ,
}

\DeclareAcronym{hf}{
   short = HF ,
   long = Hartree-Fock ,
}

\DeclareAcronym{mcscf}{
   short = MCSCF ,
   long = multi-configurational self-consistent field ,
}

\DeclareAcronym{sop}{
   short = SOP ,
   long = sum-of-products ,
}

\DeclareAcronym{midascpp}{
   short = MidasCpp ,
   long = Molecular Interactions{,} Dynamics And Simulations Chemistry Program Package ,
}

\DeclareAcronym{mpi}{
   short = MPI ,
   long = message passing interface ,
}

\DeclareAcronym{ode}{
   short = ODE ,
   long  = ordinary differential equation ,
   short-plural = s ,
   long-plural = s ,
   short-indefinite = an ,
   long-indefinite = an ,
}

\DeclareAcronym{bch}{
   short = BCH ,
   long = Baker-Campbell-Hausdorff ,
}

\DeclareAcronym{sr}{
   short = SR ,
   long = single-reference ,
}

\DeclareAcronym{mr}{
   short = MR ,
   long = multi-reference ,
}

\DeclareAcronym{dof}{
   short = DOF ,
   long = degree of freedom ,
   short-plural-form = DOFs ,
   long-plural-form = degrees of freedom ,
}

\DeclareAcronym{hp}{
   short = HP ,
   long = Hartree product ,
}

\DeclareAcronym{tdbvp}{
   short = TDBVP ,
   long  = time-dependent bivariational principle ,
   short-plural = s ,
   long-plural = s ,
   short-indefinite = a ,
   long-indefinite = a ,
}

\DeclareAcronym{dfvp}{
   short = DFVP ,
   long  = Dirac-Frenkel variational principle ,
}

\DeclareAcronym{ele}{
   short = ELE ,
   long  = Euler-Lagrange equation ,
   short-plural = s ,
   long-plural = s ,
}

\DeclareAcronym{mrcc}{
   short = MRCC ,
   long = multi-reference coupled cluster ,
}

\DeclareAcronym{tdfvci}{
   short = TDFVCI ,
   long = time-dependent full vibrational configuration interaction ,
}

\DeclareAcronym{tdfci}{
   short = TDFCI ,
   long = time-dependent full configuration interaction ,
}

\DeclareAcronym{tdevcc}{
   short = TDEVCC ,
   long  = time-dependent extended vibrational coupled cluster ,
   short-plural = s ,
   long-plural = s ,
   short-indefinite = a ,
   long-indefinite = a ,
}

\DeclareAcronym{holc}{
   short = HOLC ,
   long = hybrid optimized and localized coordinate ,
}

\DeclareAcronym{nc}{
   short = NC ,
   long = normal coordinate ,
}

\DeclareAcronym{oc}{
   short = OC ,
   long = optimized coordinate ,
}

\DeclareAcronym{lc}{
   short = LC ,
   long = localized coordinate ,
}

\DeclareAcronym{acf}{
   short = ACF ,
   long = autocorrelation function ,
}

\DeclareAcronym{fwhm}{
   short = FWHM ,
   long  = full width at half maximum ,
   short-plural = s ,
   long-plural = full widths at half maxima ,
   short-indefinite = an ,
   long-indefinite = a ,
}

\DeclareAcronym{tdmvcc}{
   short = TDMVCC ,
   long = time-dependent vibrational coupled cluster with time-dependent modals ,
}

\DeclareAcronym{vqe}{
   short = VQE ,
   long = variational quantum eigensolver ,
}

\DeclareAcronym{qwc}{
   short = QWC ,
   long = qubit-wise commutativity ,
}

\DeclareAcronym{fc}{
   short = FC ,
   long = full commutativity ,
}

\DeclareAcronym{siqwc}{
   short = SI/QWC ,
   long = sorted insertion with qubit-wise commutativity ,
}

\DeclareAcronym{sifc}{
   short = SI/FC ,
   long = sorted insertion with full commutativity ,
}

\DeclareAcronym{siqwcmcr}{
   short = SI/QWC/MCR ,
   long = sorted insertion with qubit-wise commutativity and mode-combination logic,
}

\DeclareAcronym{sifcmcr}{
   short = SI/FC/MCR ,
   long = sorted insertion with qubit-wise commutativity and mode-combination logic,
}

\DeclareAcronym{si}{
   short = SI ,
   long = sorted insertion ,
}

\DeclareAcronym{uvccsd}{
   short = UVCCSD ,
   long = unitary vibrational coupled cluster with singles and doubles ,
}

\DeclareAcronym{uccsd}{
   short = UCCSD ,
   long = unitary coupled cluster with singles and doubles ,
}

%% file: local.tex
\newcommand{\mbf}[1]{\mathbf{#1}}
\newcommand{\nn}{\nonumber\\}

\newcommand{\crea}[2]{a^{#1\mspace{-0.5mu}\raisebox{0.3ex}{$\scriptstyle\dagger$}}_{\mspace{-0mu}#2}}

\newcommand{\anni}[2]{a^{#1}_{\mspace{-0mu}#2}}

%% file: scaling.bbl
\begin{thebibliography}{10}

\bibitem{daley_practical_2022}
A.~J. Daley, I.~Bloch, C.~Kokail, S.~Flannigan, N.~Pearson, M.~Troyer, and
  P.~Zoller, ``Practical quantum advantage in quantum simulation,'' {\em
  Nature}, vol.~607, pp.~667--676, July 2022.
\newblock Number: 7920 Publisher: Nature Publishing Group.

\bibitem{lee_is_2022}
S.~Lee, J.~Lee, H.~Zhai, Y.~Tong, A.~M. Dalzell, A.~Kumar, P.~Helms, J.~Gray,
  Z.-H. Cui, W.~Liu, M.~Kastoryano, R.~Babbush, J.~Preskill, D.~R. Reichman,
  E.~T. Campbell, E.~F. Valeev, L.~Lin, and G.~K.-L. Chan, ``Is there evidence
  for exponential quantum advantage in quantum chemistry?,'' Aug. 2022.
\newblock arXiv:2208.02199 [physics, physics:quant-ph].

\bibitem{elfving_how_2020}
V.~E. Elfving, B.~W. Broer, M.~Webber, J.~Gavartin, M.~D. Halls, K.~P. Lorton,
  and A.~Bochevarov, ``How will quantum computers provide an industrially
  relevant computational advantage in quantum chemistry?,'' Sept. 2020.
\newblock arXiv:2009.12472 [physics, physics:quant-ph].

\bibitem{aspuru-guzik_simulated_2005}
A.~Aspuru-Guzik, A.~D. Dutoi, P.~J. Love, and M.~Head-Gordon, ``Simulated
  quantum computation of molecular energies,'' {\em Science (New York, N.Y.)},
  vol.~309, pp.~1704--1707, Sept. 2005.

\bibitem{anand_quantum_2022}
A.~Anand, P.~Schleich, S.~Alperin-Lea, P.~W.~K. Jensen, S.~Sim,
  M.~Díaz-Tinoco, J.~S. Kottmann, M.~Degroote, A.~F. Izmaylov, and
  A.~Aspuru-Guzik, ``A {Quantum} {Computing} {View} on {Unitary} {Coupled}
  {Cluster} {Theory},'' {\em Chemical Society Reviews}, vol.~51, no.~5,
  pp.~1659--1684, 2022.
\newblock arXiv:2109.15176 [physics, physics:quant-ph].

\bibitem{babbush_exponentially_2016}
R.~Babbush, D.~W. Berry, I.~D. Kivlichan, A.~Y. Wei, P.~J. Love, and
  A.~Aspuru-Guzik, ``Exponentially more precise quantum simulation of fermions
  in second quantization,'' {\em New Journal of Physics}, vol.~18, p.~033032,
  Mar. 2016.
\newblock Publisher: IOP Publishing.

\bibitem{babbush_exponentially_2017}
R.~Babbush, D.~W. Berry, Y.~R. Sanders, I.~D. Kivlichan, A.~Scherer, A.~Y. Wei,
  P.~J. Love, and A.~Aspuru-Guzik, ``Exponentially more precise quantum
  simulation of fermions in the configuration interaction representation,''
  {\em Quantum Science and Technology}, vol.~3, p.~015006, Dec. 2017.
\newblock Publisher: IOP Publishing.

\bibitem{mcclean_theory_2016}
J.~R. McClean, J.~Romero, R.~Babbush, and A.~Aspuru-Guzik, ``The theory of
  variational hybrid quantum-classical algorithms,'' {\em New Journal of
  Physics}, vol.~18, p.~023023, Feb. 2016.
\newblock Publisher: IOP Publishing.

\bibitem{noauthor_variational_nodate}
``A variational eigenvalue solver on a photonic quantum processor {\textbar}
  {Nature} {Communications}.''

\bibitem{sawaya_near-_2021}
N.~P.~D. Sawaya, F.~Paesani, and D.~P. Tabor, ``Near- and long-term quantum
  algorithmic approaches for vibrational spectroscopy,'' {\em Physical Review
  A}, vol.~104, p.~062419, Dec. 2021.
\newblock Publisher: American Physical Society.

\bibitem{sawaya_resource-efficient_2020}
N.~P.~D. Sawaya, T.~Menke, T.~H. Kyaw, S.~Johri, A.~Aspuru-Guzik, and G.~G.
  Guerreschi, ``Resource-efficient digital quantum simulation of d-level
  systems for photonic, vibrational, and spin-s {Hamiltonians},'' {\em npj
  Quantum Information}, vol.~6, pp.~1--13, June 2020.
\newblock Number: 1 Publisher: Nature Publishing Group.

\bibitem{mcardle_digital_2019}
S.~McArdle, A.~Mayorov, X.~Shan, S.~Benjamin, and X.~Yuan, ``Digital quantum
  simulation of molecular vibrations,'' {\em Chemical Science}, vol.~10,
  no.~22, pp.~5725--5735, 2019.
\newblock arXiv:1811.04069 [quant-ph].

\bibitem{ollitrault_hardware_2020}
P.~J. Ollitrault, A.~Baiardi, M.~Reiher, and I.~Tavernelli, ``Hardware
  {Efficient} {Quantum} {Algorithms} for {Vibrational} {Structure}
  {Calculations},'' {\em Chemical Science}, vol.~11, no.~26, pp.~6842--6855,
  2020.
\newblock arXiv:2003.12578 [quant-ph].

\bibitem{gonthier_identifying_2020}
J.~F. Gonthier, M.~D. Radin, C.~Buda, E.~J. Doskocil, C.~M. Abuan, and
  J.~Romero, ``Identifying challenges towards practical quantum advantage
  through resource estimation: the measurement roadblock in the variational
  quantum eigensolver,'' Tech. Rep. arXiv:2012.04001, arXiv, Dec. 2020.
\newblock arXiv:2012.04001 [quant-ph] type: article.

\bibitem{wang_minimizing_2021}
G.~Wang, D.~E. Koh, P.~D. Johnson, and Y.~Cao, ``Minimizing estimation runtime
  on noisy quantum computers,'' {\em PRX Quantum}, vol.~2, p.~010346, Mar.
  2021.
\newblock arXiv:2006.09350 [quant-ph].

\bibitem{zhao_measurement_2020}
A.~Zhao, A.~Tranter, W.~M. Kirby, S.~F. Ung, A.~Miyake, and P.~Love,
  ``Measurement reduction in variational quantum algorithms,'' {\em Physical
  Review A}, vol.~101, p.~062322, June 2020.
\newblock arXiv:1908.08067 [quant-ph].

\bibitem{verteletskyi_measurement_2020}
V.~Verteletskyi, T.-C. Yen, and A.~F. Izmaylov, ``Measurement {Optimization} in
  the {Variational} {Quantum} {Eigensolver} {Using} a {Minimum} {Clique}
  {Cover},'' {\em The Journal of Chemical Physics}, vol.~152, p.~124114, Mar.
  2020.
\newblock arXiv:1907.03358 [physics, physics:quant-ph].

\bibitem{yen_deterministic_2022}
T.-C. Yen, A.~Ganeshram, and A.~F. Izmaylov, ``Deterministic improvements of
  quantum measurements with grouping of compatible operators, non-local
  transformations, and covariance estimates,'' Apr. 2022.
\newblock arXiv:2201.01471 [physics, physics:quant-ph].

\bibitem{bansingh_fidelity_2022}
Z.~P. Bansingh, T.-C. Yen, P.~D. Johnson, and A.~F. Izmaylov, ``Fidelity
  overhead for non-local measurements in variational quantum algorithms,'' May
  2022.
\newblock arXiv:2205.07113 [physics, physics:quant-ph].

\bibitem{yen_measuring_2020}
T.-C. Yen, V.~Verteletskyi, and A.~F. Izmaylov, ``Measuring all compatible
  operators in one series of a single-qubit measurements using unitary
  transformations,'' Tech. Rep. arXiv:1907.09386, arXiv, Mar. 2020.
\newblock arXiv:1907.09386 [physics, physics:quant-ph] type: article.

\bibitem{hamamura_efficient_2020}
I.~Hamamura and T.~Imamichi, ``Efficient evaluation of quantum observables
  using entangled measurements,'' {\em npj Quantum Information}, vol.~6,
  pp.~1--8, June 2020.
\newblock Number: 1 Publisher: Nature Publishing Group.

\bibitem{crawford_efficient_2021}
O.~Crawford, B.~v. Straaten, D.~Wang, T.~Parks, E.~Campbell, and S.~Brierley,
  ``Efficient quantum measurement of {Pauli} operators in the presence of
  finite sampling error,'' {\em Quantum}, vol.~5, p.~385, Jan. 2021.
\newblock Publisher: Verein zur Förderung des Open Access Publizierens in den
  Quantenwissenschaften.

\bibitem{choi_improving_2022}
S.~Choi, T.-C. Yen, and A.~F. Izmaylov, ``Improving quantum measurements by
  introducing "ghost" {Pauli} products,'' Aug. 2022.
\newblock arXiv:2208.06563 [physics, physics:quant-ph].

\bibitem{izmaylov_revising_2019}
A.~F. Izmaylov, T.-C. Yen, and I.~G. Ryabinkin, ``Revising the measurement
  process in the variational quantum eigensolver: is it possible to reduce the
  number of separately measured operators?,'' {\em Chemical Science}, vol.~10,
  pp.~3746--3755, Mar. 2019.
\newblock Publisher: The Royal Society of Chemistry.

\bibitem{izmaylov_unitary_2019}
A.~F. Izmaylov, T.-C. Yen, R.~A. Lang, and V.~Verteletskyi, ``Unitary
  partitioning approach to the measurement problem in the {Variational}
  {Quantum} {Eigensolver} method,'' Tech. Rep. arXiv:1907.09040, arXiv, Oct.
  2019.
\newblock arXiv:1907.09040 [physics, physics:quant-ph] type: article.

\bibitem{hadfield_measurements_2020}
C.~Hadfield, S.~Bravyi, R.~Raymond, and A.~Mezzacapo, ``Measurements of
  {Quantum} {Hamiltonians} with {Locally}-{Biased} {Classical} {Shadows},''
  June 2020.
\newblock arXiv:2006.15788 [quant-ph].

\bibitem{huang_efficient_2021}
H.-Y. Huang, R.~Kueng, and J.~Preskill, ``Efficient {Estimation} of {Pauli}
  {Observables} by {Derandomization},'' {\em Physical Review Letters},
  vol.~127, p.~030503, July 2021.
\newblock Publisher: American Physical Society.

\bibitem{hadfield_adaptive_2021}
C.~Hadfield, ``Adaptive {Pauli} {Shadows} for {Energy} {Estimation},'' May
  2021.
\newblock arXiv:2105.12207 [quant-ph].

\bibitem{wu_overlapped_2021}
B.~Wu, J.~Sun, Q.~Huang, and X.~Yuan, ``Overlapped grouping measurement: {A}
  unified framework for measuring quantum states,'' May 2021.
\newblock arXiv:2105.13091 [quant-ph].

\bibitem{huggins_efficient_2021}
W.~J. Huggins, J.~McClean, N.~Rubin, Z.~Jiang, N.~Wiebe, K.~B. Whaley, and
  R.~Babbush, ``Efficient and {Noise} {Resilient} {Measurements} for {Quantum}
  {Chemistry} on {Near}-{Term} {Quantum} {Computers},'' {\em npj Quantum
  Information}, vol.~7, p.~23, Dec. 2021.
\newblock arXiv:1907.13117 [physics, physics:quant-ph].

\bibitem{yen_cartan_2021}
T.-C. Yen and A.~F. Izmaylov, ``Cartan {Subalgebra} {Approach} to {Efficient}
  {Measurements} of {Quantum} {Observables},'' {\em PRX Quantum}, vol.~2,
  p.~040320, Oct. 2021.
\newblock Publisher: American Physical Society.

\bibitem{jena_pauli_2019}
A.~Jena, S.~Genin, and M.~Mosca, ``Pauli {Partitioning} with {Respect} to
  {Gate} {Sets},'' July 2019.
\newblock arXiv:1907.07859 [quant-ph].

\bibitem{watsonSimplificationMolecularVibrationrotation1968}
J.~K. Watson, ``Simplification of the molecular vibration-rotation
  hamiltonian,'' {\em Molecular Physics}, vol.~15, pp.~479--490, Jan. 1968.

\bibitem{watsonVibrationrotationHamiltonianLinear1970}
J.~K. Watson, ``The vibration-rotation hamiltonian of linear molecules,'' {\em
  Molecular Physics}, vol.~19, pp.~465--487, Oct. 1970.

\bibitem{sparta_adaptive_2009}
M.~Sparta, D.~Toffoli, and O.~Christiansen, ``An adaptive density-guided
  approach for the generation of potential energy surfaces of polyatomic
  molecules,'' {\em Theoretical Chemistry Accounts}, vol.~123, pp.~413--429,
  Aug. 2009.

\bibitem{yagi_optimized_2012}
K.~Yagi, M.~Keçeli, and S.~Hirata, ``Optimized coordinates for anharmonic
  vibrational structure theories,'' {\em The Journal of Chemical Physics},
  vol.~137, p.~204118, Nov. 2012.
\newblock Publisher: American Institute of Physics.

\bibitem{thomsen_optimized_2014}
B.~Thomsen, K.~Yagi, and O.~Christiansen, ``Optimized coordinates in
  vibrational coupled cluster calculations,'' {\em The Journal of Chemical
  Physics}, vol.~140, p.~154102, Apr. 2014.
\newblock Publisher: American Institute of Physics.

\bibitem{jacob_localizing_2009}
C.~R. Jacob and M.~Reiher, ``Localizing normal modes in large molecules,'' {\em
  The Journal of Chemical Physics}, vol.~130, p.~084106, Feb. 2009.

\bibitem{klinting_hybrid_2015}
E.~L. Klinting, C.~König, and O.~Christiansen, ``Hybrid {Optimized} and
  {Localized} {Vibrational} {Coordinates},'' {\em The Journal of Physical
  Chemistry A}, vol.~119, pp.~11007--11021, Nov. 2015.
\newblock Publisher: American Chemical Society.

\bibitem{hattigCommunicationsAccurateEfficient2010}
C.~H{\"a}ttig, D.~P. Tew, and A.~K{\"o}hn, ``Communications: {{Accurate}} and
  efficient approximations to explicitly correlated coupled-cluster singles and
  doubles, {{CCSD-F12}},'' {\em The Journal of Chemical Physics}, vol.~132,
  p.~231102, June 2010.

\bibitem{petersonSystematicallyConvergentBasis2008}
K.~A. Peterson, T.~B. Adler, and H.-J. Werner, ``Systematically convergent
  basis sets for explicitly correlated wavefunctions: {{The}} atoms {{H}},
  {{He}}, {{B}}\textendash{{Ne}}, and {{Al}}\textendash{{Ar}},'' {\em The
  Journal of Chemical Physics}, vol.~128, p.~084102, Feb. 2008.

\bibitem{TurbomoleV7}
``Turbomole {{V7}}.5.'' TURBOMOLE GmbH.

\bibitem{artiukhinMidasCpp2022}
D.~G. Artiukhin, O.~Christiansen, I.~H. Godtliebsen, E.~M. Gras, W.~Gy{\H
  o}rffy, M.~B. Hansen, M.~B. Hansen, E.~L. Klinting, J.~Kongsted,
  C.~K{\"o}nig, D.~Madsen, N.~K. Madsen, K.~Monrad, G.~Schmitz, P.~Seidler,
  K.~Sneskov, M.~Sparta, B.~Thomsen, D.~Toffoli, A.~Zoccante, M.~G. H{\o}jlund,
  N.~M. H{\o}yer, and A.~B. Jensen, ``{{MidasCpp}},'' 2022.

\end{thebibliography}
